\documentclass[11pt]{article}
\usepackage{amsmath,amssymb}
\usepackage[letterpaper,left=0.82in,right=0.82in,top=0.82in,bottom=0.82in]{geometry}
\usepackage{microtype}
\usepackage{orcidlink}
\usepackage{hyperref}
\hypersetup{colorlinks=true, linkcolor=black, citecolor=black, urlcolor=blue}
\usepackage{setspace}
\usepackage{indentfirst}
\RequirePackage[numbers,sort&compress]{natbib}
\begin{document}
\onehalfspacing

\begin{center}
\LARGE{Singularity-Rank Signatures in Generalized Dirac Oscillators near a BTZ Horizon}
\par\end{center}

\vspace{0.3cm}

\begin{center}
    {\bf João V. R. Alencar}$^{*}$\orcidlink{0009-0008-3042-5807}\\
    {\it Department of Teleinformatics Engineering, Federal University of Cear\'{a} (UFC), Fortaleza, CE, 60440-900, Brazil}\\
    {\small $^{*}$Corresponding author: \href{mailto:joaoalencar@alu.ufc.br}{joaoalencar@alu.ufc.br}}\\
    \vspace{0.1cm}

    {\bf Allan R. P. Moreira\orcidlink{0000-0002-6535-493X}}\\ 
    {\it Secretaria da Educa\c{c}\~{a}o do Cear\'{a} (SEDUC), Coordenadoria Regional de Desenvolvimento da Educa\c{c}\~{a}o (CREDE 9), Horizonte, Cear\'{a}, 62880-384, Brazil}\\
    {\it Postgraduate Program in Electrical and Computer Engineering, Federal University of Cear\'{a}, Sobral, Cear\'{a}, 62010-560, Brazil}\\
    {\small \href{mailto:allan.moreira@fisica.ufc.br}{allan.moreira@fisica.ufc.br}}\\
    \vspace{0.1cm}

    {\bf João B. R. Silva\orcidlink{0000-0002-4004-0926}}\\
    {\it Department of Teleinformatics Engineering, Federal University of Cear\'{a} (UFC), Fortaleza, CE, 60440-900, Brazil}\\
    {\small \href{mailto:joaobrs@ufc.br}{joaobrs@ufc.br}}\\
    \vspace{0.1cm}

 {\bf Abdullah Guvendi}\orcidlink{0000-0003-0564-9899}\\
    {\it Department of Basic Sciences, Faculty of Science, Erzurum Technical University, Erzurum, 25050, T\"{u}rkiye}\\
    {\small \href{mailto:abdullah.guvendi@erzurum.edu.tr}{abdullah.guvendi@erzurum.edu.tr}}\\

\vspace{0.4cm}

\end{center}

\begin{abstract}
We study singular generalized Dirac-oscillator couplings in the near-horizon region of a nonextremal BTZ black hole. An effective covariant construction adapted to the stationary horizon congruence relates the two singular profiles to the proper acceleration and its radial gradient, whose leading near-horizon behaviors are $1/\rho$ and $1/\rho^2$. This motivates the family $\mathcal{L}_p(\rho)=\lambda_0+\beta_p/\rho^p$, with $p=1,2$. The $p=1$ system has a regular singular point and is Fuchsian, whereas $p=2$ is the first irregular case; more generally, a leading $\rho^{-p}$ interaction with $p>1$ has Poincar\'e rank $p-1$. For $p=1$, a constant spinor transformation reduces the matrix-Coulomb system exactly to the Whittaker equation, while elimination in the original basis gives an equivalent confluent-Heun representation with an apparent finite singularity. For $p=2$, the local branches contain the essential factors $e^{\pm\beta_2/\rho}\rho^{-1/2}$, and their coefficient expansions exhibit generic factorial late-order growth. The finite-radius response $\mathcal{Z}_p(\Omega;\rho_0)=\psi_{2,p}(\rho_0)/\psi_{1,p}(\rho_0)$ obeys an exact Riccati equation whose large-damping expansion first becomes sensitive to the radial gradient of the interaction at order $\gamma^{-3}$. Even when the two couplings are matched at $\rho_0$, this response retains information about their different pole orders. The response functions provide local spinorial data for subsequent matching to the complete BTZ exterior.
\end{abstract}
\vspace{0.3cm}

\section{Introduction}
\label{sec:introduction}

Spinorial fields on black-hole backgrounds probe the simultaneous influence of horizon geometry, relativistic interactions, and boundary conditions. The Ba\~nados--Teitelboim--Zanelli (BTZ) black hole is a particularly controlled arena: it is an exact solution of $(2+1)$-dimensional gravity with negative cosmological constant, and many field equations remain analytically tractable \cite{btz1992,banados1993,carlip1995,carlip2005}. Early quasinormal-mode studies established close connections among AdS black-hole perturbations, BTZ spectra, thermal relaxation, and the dual conformal field theory \cite{horowitz2000,cardoso2001btz,birmingham2002}. More recent work has extended this program through ladder constructions, anyonic perturbations, charged-field instabilities, and mixed-boundary-condition superradiance \cite{katagiri2022,anyons2024,fontana2024,fontana2024instability,konewko2024}. These developments complement general black-hole spectroscopy and spectral-method reviews \cite{berti2009,konoplya2011,hatsuda2021}.

Fermionic propagation provides a distinct sector because its radial dynamics is first order and its admissible boundary data are constrained by the Dirac current. Exact studies of BTZ fermions have addressed quantum properties, torsionful geometries, and noncommutative dual descriptions \cite{belgiorno2010,becar2014,gupta2017}, while near-horizon analyses have recently treated a one-body Dirac oscillator and interacting fermion--antifermion systems, including a Cornell-type generalized Dirac-oscillator coupling \cite{guvendi2024,guvendi2024pair,guvendimustafa2025}. The present work starts from the separated one-body system of Ref.~\cite{guvendi2024} and asks how singular generalized-oscillator couplings change its analytical class and local response.

The Dirac oscillator, introduced by Moshinsky and Szczepaniak through a linear nonminimal momentum substitution \cite{moshinsky1989}, is a standard exactly solvable model of relativistic confinement and spin--orbit coupling; an overview of its structure and applications is given in Ref.~\cite{sadurni2011}. Geometric and deformed-background realizations have been studied in cosmic-string spacetimes, anti-de Sitter kinematics, and spinning defect geometries \cite{carvalho2011,hamil2018,cunha2020}. Generalized Dirac oscillators replace the linear spatial profile by a more general function and include Cornell and singular profiles in curved backgrounds \cite{deng2018}; a Cornell-type generalized coupling has also been used in the near-horizon BTZ two-body problem \cite{guvendi2024pair}. Recent extensions further include Aharonov--Casher coupling in a cosmic-string spacetime \cite{chen2022}, unitary mappings between flat- and curved-space Dirac problems with shape-invariant potentials \cite{oliveira2022}, and an exact Whittaker treatment in a cloud-of-strings geometry \cite{oliveira2026}. This literature shows that the coupling prescription, spin connection, and chosen spinor basis jointly determine the solvable radial structure.

Singular oscillator extensions provide the second ingredient. The isotonic oscillator is naturally defined on the half-line, and its superpotential combines linear and inverse-coordinate terms. Its nonrelativistic and relativistic generalizations exhibit exact or quasi-exact sectors \cite{goldman1961,agboola2012,agboola2014}; the recent Dirac-isotonic construction in $(1+1)$ and $(2+1)$ dimensions makes this connection explicit \cite{ghosh2025}. Because the proper near-horizon coordinate obeys $\rho>0$, the BTZ radial problem supplies a natural domain in which to examine singular isotonic-inspired couplings. Below, this half-line motivation is combined with geometric scales associated with the stationary horizon congruence.

We organize the two deformations as the family
\begin{equation}
    \mathcal{L}_p(\rho)
    =
    \lambda_0+\frac{\beta_p}{\rho^p},
    \qquad
    p\in\{1,2\},
    \label{eq:intro_lambda_iso}
\end{equation}
where $\lambda_0$ is the constant oscillator-like term after the near-horizon rescaling. We show below that these two singular profiles arise as the leading near-horizon terms of an effective covariant generalized-oscillator interaction built from the proper acceleration of the horizon-generating congruence. The interaction is defined on the fixed BTZ background and does not modify the geometry.

The two powers are structurally distinguished already at the first-order level. The $p=1$ system has only a simple pole at the horizon and is therefore Fuchsian; it is the regular-singular threshold of the family. The $p=2$ system has a pole of order two and is the first irregular member. For $p=1$, the complete first-order problem belongs to the matrix-Coulomb class and is exactly reducible to Whittaker functions by a constant spinor transformation. Eliminating one component in the original basis instead produces a confluent-Heun equation with a finite apparent singularity, a distinction relevant to the interpretation of singular points in black-hole perturbation theory \cite{bonelli2022,minuccimacedo2025}. For $p=2$, the local spinor contains essential factors $e^{\pm\beta_2/\rho}\rho^{-1/2}$, and the remaining formal series generically develops factorial late-order growth.

We compare the two radial problems on the same footing. For $p=1$, we obtain the full exact spinor and demonstrate the equivalence of its Whittaker and componentwise Heun representations. For $p=2$, we determine both essential branches and the coefficient recurrences, including the sign-dependent choice of the locally bounded branch. Building on the finite-radius near-horizon Dirac-oscillator framework of Ref.~\cite{guvendi2024}, the two models are compared through the finite-radius spinor response
\begin{equation}
    \mathcal{Z}_p(\Omega;\rho_0)
    =
    \frac{\psi_{2,p}(\rho_0;\Omega)}
         {\psi_{1,p}(\rho_0;\Omega)}.
    \label{eq:spinor_response_definition}
\end{equation}
This ratio supplies matching data at the outer edge of the near-horizon region; by itself, it is not a quasinormal-mode condition. Its Riccati equation also yields a controlled large-damping expansion that identifies which terms are universal and which retain information about the radial interaction. A global spectrum additionally requires propagation through the complete exterior and an admissible asymptotic AdS boundary condition \cite{ishibashi2004,wang2018,wang2020,deoliveira2022bc}; Robin boundary conditions in BTZ and BTZ-analogue settings have also been studied in Refs.~\cite{bussola2017,dappiaggi2018,deoliveira2023robin}. We therefore treat the solutions of $\operatorname{Re}\mathcal{Z}_p=\eta$ as projected local-response intersections and separately track the radial flux.

The remainder of this paper is organized as follows. Section~\ref{sec:btz_dirac_system} derives the separated near-horizon spinor system and introduces the family \eqref{eq:intro_lambda_iso}. Sections~\ref{sec:p1_isotonic} and \ref{sec:p2_isotonic} analyze the $p=1$ and $p=2$ models, respectively. Section~\ref{sec:finite_radius_response} develops the finite-radius spinor response and its analytical large-damping expansion. The componentwise confluent-Heun representation, its analytical equivalence to the first-order spinor formulation, and the constant-coupling limit are collected in Appendix~\ref{sec:heun_reconstruction}.

Throughout this work, we use natural units, $\hbar=c=1$. Accordingly, $m$, $\lambda_0$, and $\mathcal{L}_p(\rho)$ have dimensions of inverse length. Since $\rho$ has dimensions of length, $[\beta_p]=L^{p-1}$; hence $\beta_1$ is dimensionless and $\beta_2$ has dimensions of length.

\section{Near-horizon BTZ geometry and radial Dirac system}
\label{sec:btz_dirac_system}

\subsection{BTZ background and near-horizon limit}
\label{subsec:btz_near_horizon}

The BTZ black hole is a solution of $(2+1)$-dimensional gravity with negative cosmological constant
\begin{equation}
    \Lambda_{\rm cc}
    =
    -\frac{1}{\ell^2},
\end{equation}
where $\ell$ is the anti-de Sitter curvature radius. Using the signature convention $(+,-,-)$, the rotating BTZ line element is written as \cite{btz1992,banados1993}
\begin{equation}
    ds^2
    =
    N^2(r)\,dt^2
    -
    \frac{dr^2}{N^2(r)}
    -
    r^2\left[d\phi+N^\phi(r)\,dt\right]^2 ,
    \label{eq:btz_metric}
\end{equation}
with
\begin{equation}
    N^2(r)
    =
    -M+\frac{r^2}{\ell^2}
    +
    \frac{J^2}{4r^2},
    \qquad
    N^\phi(r)
    =
    -\frac{J}{2r^2}.
    \label{eq:btz_lapse_shift}
\end{equation}
The parameters $M$ and $J$ denote the mass and angular momentum of the black hole. We choose the angular orientation such that $J\geq0$; the case $J<0$ is obtained by reversing $\phi$. For a nonextremal BTZ black hole, $M\ell>|J|$, the lapse function has two positive roots, $r_+$ and $r_-$, corresponding to the outer and inner horizons. In terms of these radii \cite{btz1992,banados1993},
\begin{equation}
    M
    =
    \frac{r_+^2+r_-^2}{\ell^2},
    \qquad
    J
    =
    \frac{2r_+r_-}{\ell},
    \label{eq:btz_mass_angular_momentum}
\end{equation}
and the lapse can be written as
\begin{equation}
    N^2(r)
    =
    \frac{(r^2-r_+^2)(r^2-r_-^2)}
    {\ell^2 r^2}.
    \label{eq:btz_lapse_horizon}
\end{equation}

The near-horizon geometry is most naturally described in a frame corotating with the outer horizon. The horizon angular velocity is
\begin{equation}
    \Omega_H
    =
    -N^\phi(r_+)
    =
    \frac{r_-}{\ell r_+}.
    \label{eq:horizon_angular_velocity}
\end{equation}
The corresponding horizon-generating Killing vector is \cite{banados1993,carlip1995}
\begin{equation}
    \chi=\partial_t+\Omega_H\partial_\phi ,
\end{equation}
which becomes null at $r=r_+$. We then introduce the corotating angular coordinate
\begin{equation}
    \varphi
    =
    \phi-\Omega_H t.
    \label{eq:corotating_angle}
\end{equation}
The surface gravity of the nonextremal rotating BTZ black hole is \cite{banados1993,carlip1995}
\begin{equation}
    \kappa
    =
    \frac{r_+^2-r_-^2}{\ell^2 r_+}.
    \label{eq:btz_surface_gravity}
\end{equation}
For a nonextremal black hole, $\kappa>0$.
The extremal limit $r_+=r_-$ is excluded because $\kappa\rightarrow0$ and the Rindler-type near-horizon expansion used below is no longer valid.

Let $\rho$ denote the proper radial distance from the outer horizon, with $\rho=0$ at $r=r_+$.
Since
\begin{equation}
    \rho=\int_{r_+}^{r}\frac{dr'}{N(r')}.
\end{equation}
the near-horizon relation between the radial coordinate and the proper distance is
\begin{equation}
    r-r_+
    =
    \frac{\kappa}{2}\rho^2+O(\rho^4).
\end{equation}
Near the horizon, the lapse behaves as
\begin{equation}
    N(r)
    =
    \kappa\rho
    +
    O(\rho^3),
    \label{eq:lapse_near_horizon}
\end{equation}
and the metric reduces, to leading order in $\rho$, to
\begin{equation}
    ds^2
    =
    \kappa^2\rho^2\,dt^2
    -
    d\rho^2
    -
    R_+^2\,d\varphi^2
    +O(\rho^2),
    \qquad
    R_+\equiv r_+ .
    \label{eq:near_horizon_metric}
\end{equation}
The coordinate $\rho$ is therefore the proper radial distance from the horizon, while $R_+$ is the circumference radius of the angular direction at the horizon.

The near-horizon approximation is valid on a finite radial interval
\begin{equation}
    0<\rho\leq\rho_0,
    \qquad
    \rho_0\ll\ell,\ r_+,
    \label{eq:near_horizon_domain}
\end{equation}
where $\rho_0$ denotes the outer scale at which the local near-horizon description is truncated. The horizon is located at the endpoint $\rho=0$, and the radial coordinate is naturally restricted to the half-line $\rho>0$. This property will be important when inverse-radial isotonic deformations are introduced.

For the static BTZ black hole, $r_-=0$, one has
\begin{equation}
    \kappa=\frac{r_+}{\ell^2}.
\end{equation}
In that special case, defining $\alpha=r_+/\ell$ gives $\kappa=\alpha/\ell$ and $R_+=\alpha\ell$,
which allows the near-horizon parameters to be expressed in terms of the single dimensionless ratio $\alpha=r_+/\ell$.
In the rotating case, however, $\kappa$ and $R_+$ are independent near-horizon scales. We keep them distinct in order to avoid imposing a static one-scale parametrization on the generic nonextremal geometry.

\subsection{Stationary horizon congruence and geometric interaction scales}
\label{subsec:stationary_horizon_scales}

The leading Rindler form \eqref{eq:near_horizon_metric} is sufficient for the spinor analysis below, but the full BTZ geometry provides a useful geometric interpretation of the singular radial profiles. In the exterior region where the horizon generator $\chi$ is timelike, following the standard normalization of the horizon-generating Killing field for stationary observers \cite{carlip1995,brown1994}, we define the normalized stationary congruence
\begin{equation}
    u^\mu
    =
    \frac{\chi^\mu}{\sqrt{\chi^2}} .
    \label{eq:stationary_congruence}
\end{equation}
Using Eqs.~\eqref{eq:btz_metric}--\eqref{eq:horizon_angular_velocity}, its norm is
\begin{equation}
    \chi^2
    =
    \frac{
        (r^2-r_+^2)(r_+^2-r_-^2)
    }{
        \ell^2 r_+^2
    } .
    \label{eq:chi_norm_btz}
\end{equation}
For the proper-distance coordinate introduced above, the exact radial relation is
\begin{equation}
    r^2
    =
    r_+^2
    +
    (r_+^2-r_-^2)
    \sinh^2\!\left(\frac{\rho}{\ell}\right),
    \label{eq:r_of_rho_exact}
\end{equation}
whose small-$\rho$ expansion reproduces Eq.~\eqref{eq:lapse_near_horizon} and the relation $r-r_+=\kappa\rho^2/2+O(\rho^4)$.

Let
\begin{equation}
    a^\mu
    =
    u^\nu\nabla_\nu u^\mu,
    \qquad
    \mathfrak{a}
    =
    \sqrt{-a_\mu a^\mu}
    \label{eq:proper_acceleration_definition}
\end{equation}
be the proper acceleration of this stationary congruence. Direct evaluation in the full rotating BTZ geometry gives
\begin{equation}
    \mathfrak{a}(r)
    =
    \frac{
        \sqrt{r^2-r_-^2}
    }{
        \ell\sqrt{r^2-r_+^2}
    },
    \qquad
    \mathfrak{a}(\rho)
    =
    \frac{1}{\ell}
    \coth\!\left(\frac{\rho}{\ell}\right).
    \label{eq:btz_stationary_acceleration}
\end{equation}
Thus the leading stationary acceleration scale is universal near any nonextremal BTZ horizon,
\begin{equation}
    \mathfrak{a}(\rho)
    =
    \frac{1}{\rho}
    +
    \frac{\rho}{3\ell^2}
    +
    O(\rho^3).
    \label{eq:acceleration_near_horizon}
\end{equation}

The same congruence also carries the Tolman-redshifted Hawking temperature. With $T_H=\kappa/(2\pi)$, one finds \cite{brown1994,deserlevin1997}
\begin{equation}
    T_{\rm loc}(\rho)
    =
    \frac{T_H}{\sqrt{\chi^2}}
    =
    \frac{1}{
        2\pi\ell\sinh(\rho/\ell)
    } .
    \label{eq:local_tolman_temperature}
\end{equation}
The acceleration obeys the exact identity
\begin{equation}
    -\mathfrak{a}'(\rho)
    =
    \mathfrak{a}^2(\rho)-\frac{1}{\ell^2}
    =
    \left(2\pi T_{\rm loc}\right)^2
    =
    \frac{1}{
        \ell^2\sinh^2(\rho/\ell)
    } .
    \label{eq:acceleration_gradient_identity}
\end{equation}
Consequently, two natural stationary near-horizon profiles are
\begin{equation}
    F_1(\rho)=\mathfrak{a}(\rho),
    \qquad
    F_2(\rho)=-\mathfrak{a}'(\rho),
    \label{eq:geometric_profiles}
\end{equation}
with
\begin{equation}
    F_1(\rho)
    =
    \frac{1}{\rho}
    +
    O(\rho),
    \qquad
    F_2(\rho)
    =
    \frac{1}{\rho^2}
    -
    \frac{1}{3\ell^2}
    +
    O(\rho^2).
    \label{eq:geometric_profiles_expansion}
\end{equation}
The divergence of these stationary-frame scales is not a curvature singularity of the BTZ geometry; it reflects the increasing acceleration and Tolman temperature required to remain stationary as the horizon is approached. We use Eqs.~\eqref{eq:geometric_profiles} and \eqref{eq:geometric_profiles_expansion} to motivate the effective generalized-oscillator interaction introduced below. These stationary-frame scales do not, by themselves, provide a freely falling regularity condition.

\subsection{Dirac equation and spinorial separation}
\label{subsec:dirac_spinorial_separation}

A convenient orthonormal coframe for the near-horizon metric
\eqref{eq:near_horizon_metric} is
\begin{equation}
    e^0
    =
    \kappa\rho\,dt,
    \qquad
    e^1
    =
    d\rho,
    \qquad
    e^2
    =
    R_+\,d\varphi .
    \label{eq:dreibein}
\end{equation}
The corresponding inverse frame is
\begin{equation}
    e^\mu{}_{0}\partial_\mu
    =
    \frac{1}{\kappa\rho}\partial_t,
    \qquad
    e^\mu{}_{1}\partial_\mu
    =
    \partial_\rho,
    \qquad
    e^\mu{}_{2}\partial_\mu
    =
    \frac{1}{R_+}\partial_\varphi .
    \label{eq:inverse_dreibein}
\end{equation}
The local Lorentz metric is taken as
\begin{equation}
    \eta_{ab}
    =
    \mathrm{diag}(1,-1,-1),
\end{equation}
so that
\begin{equation}
    g_{\mu\nu}
    =
    \eta_{ab}e^a{}_\mu e^b{}_\nu .
\end{equation}

The torsion-free Cartan equation,
\begin{equation}
    de^a+\omega^a{}_{b}\wedge e^b=0,
    \label{eq:cartan_torsion_free}
\end{equation}
gives
\begin{equation}
    \omega^0{}_{1}
    =
    \kappa\,dt .
    \label{eq:spin_connection}
\end{equation}
With both Lorentz indices lowered, the connection is antisymmetric,
$\omega_{ab}=-\omega_{ba}$. Thus $\omega_{01}=\kappa\,dt$ and
$\omega_{10}=-\kappa\,dt$; equivalently, for mixed indices in the
$(+,-,-)$ convention, $\omega^1{}_{0}=\omega^0{}_{1}=\kappa\,dt$.
This contribution is responsible for the characteristic spinorial term proportional to $1/(2\rho)$ in the radial equations.

The curved-space Dirac equation for a spin-$1/2$ field of mass $m$ is
\begin{equation}
    \left[
        i\gamma^\mu
        \left(
            \partial_\mu+\Gamma_\mu
        \right)
        -m
    \right]\Psi
    =
    0,
    \label{eq:curved_dirac_equation}
\end{equation}
where
\begin{equation}
    \gamma^\mu
    =
    e^\mu{}_{a}\gamma^a,
    \qquad
    \Gamma_\mu
    =
    \frac{1}{4}
    \omega_{\mu ab}\gamma^a\gamma^b .
    \label{eq:curved_gamma_connection}
\end{equation}

Equivalently, because $\omega_{\mu ab}=-\omega_{\mu ba}$, the spinor connection can be written in the standard covariant form
\begin{equation}
    \Gamma_\mu
    =
    \frac18
    \omega_{\mu ab}
    [\gamma^a,\gamma^b].
\end{equation}
The flat gamma matrices satisfy
\begin{equation}
    \{\gamma^a,\gamma^b\}
    =
    2\eta^{ab}.
\end{equation}
We use the representation
\begin{equation}
    \gamma^0
    =
    \sigma^3,
    \qquad
    \gamma^1
    =
    i\sigma^1,
    \qquad
    \gamma^2
    =
    i\sigma^2,
    \label{eq:gamma_representation}
\end{equation}
where $\sigma^i$ are the Pauli matrices. The curved gamma matrices are then
\begin{equation}
    \gamma^t
    =
    \frac{1}{\kappa\rho}\gamma^0,
    \qquad
    \gamma^\rho
    =
    \gamma^1,
    \qquad
    \gamma^\varphi
    =
    \frac{1}{R_+}\gamma^2 .
    \label{eq:curved_gamma_matrices}
\end{equation}

Stationarity and axial symmetry allow the spinor field to be separated as
\begin{equation}
    \Psi(t,\rho,\varphi)
    =
    e^{-i\omega t}
    e^{is\varphi}
    \begin{pmatrix}
        \psi_1(\rho)\\
        \psi_2(\rho)
    \end{pmatrix},
    \label{eq:spinor_ansatz}
\end{equation}
where $\omega$ is the frequency conjugate to the corotating time coordinate and $s$ is the angular quantum number. We choose the antiperiodic spin structure around the angular circle, for which
\begin{equation}
    s
    =
    \pm\frac{1}{2},
    \pm\frac{3}{2},
    \ldots .
\end{equation}
For the periodic spin structure one instead has $s\in\mathbb Z$. The spin structure is therefore part of the global specification of the field and should not be inferred from the local radial equation alone.

Because $\varphi=\phi-\Omega_Ht$, the frequency conjugate to the original BTZ angular coordinate is
\begin{equation}
    \omega_{\rm BL}
    =
    \omega+s\Omega_H .
    \label{eq:frequency_frame_relation}
\end{equation}
Thus, a purely imaginary $\omega$ in the corotating frame has
$\operatorname{Re}\omega_{\rm BL}=s\Omega_H$ for a rotating black hole.
It is useful to introduce the dimensionless frequency parameter
\begin{equation}
    \Omega
    =
    \frac{\omega}{\kappa},
    \label{eq:dimensionless_frequency}
\end{equation}
so that the time dependence contributes to the radial equations through
\begin{equation}
    \epsilon(\rho)
    =
    \frac{\Omega}{\rho}.
    \label{eq:epsilon_definition}
\end{equation}

Before introducing the effective radial coupling, direct substitution of the separated ansatz into the geometric Dirac operator gives, for the radial spinor $\psi(\rho)=(\psi_1,\psi_2)^T$,
\begin{equation}
\left[
\gamma^0\epsilon(\rho)
+i\gamma^1
\left(
\frac{d}{d\rho}
+\frac{1}{2\rho}
\right)
-\gamma^2\frac{s}{R_+}
-m
\right]\psi(\rho)=0 .
\label{eq:intermediate_separated_dirac}
\end{equation}
The term $1/(2\rho)$ originates from the spin connection, whereas the angular term follows from $\partial_\varphi\to i s$ in the separated ansatz.

After the oscillator interaction is included and collected with the angular contribution into the effective coupling $\mathcal{L}(\rho)$, the separated equation takes the generic first-order radial form
\begin{equation}
    \left[
        \epsilon(\rho)-m
    \right]\psi_1
    +
    \left[
        \mathcal{L}(\rho)
        -
        \frac{d}{d\rho}
        -
        \frac{1}{2\rho}
    \right]\psi_2
    =
    0,
    \label{eq:generic_radial_system_1}
\end{equation}
\begin{equation}
    \left[
        \epsilon(\rho)+m
    \right]\psi_2
    +
    \left[
        \mathcal{L}(\rho)
        +
        \frac{d}{d\rho}
        +
        \frac{1}{2\rho}
    \right]\psi_1
    =
    0 .
    \label{eq:generic_radial_system_2}
\end{equation}
The function $\mathcal{L}(\rho)$ denotes the radial coupling that remains after the oscillator interaction and the angular dependence have been incorporated into the separated spinorial system.

In particular, $\mathcal{L}(\rho)$ includes both the oscillator contribution and the angular spinorial contribution proportional to $s/R_+$.

Equations~\eqref{eq:generic_radial_system_1} and
\eqref{eq:generic_radial_system_2} constitute the fundamental radial system used throughout the analysis. The second-order equations are derived only after a specific choice of $\mathcal{L}(\rho)$ has been made.

\subsection{Generalized radial coupling}
\label{subsec:generic_radial_coupling}

The near-horizon radial problem is specified by the choice of the coupling $\mathcal{L}(\rho)$. In the reference Dirac-oscillator problem, the nonminimal oscillator interaction gives a constant coupling,
\begin{equation}
    \mathcal{L}(\rho)
    =
    \lambda,
    \label{eq:usual_coupling_constant}
\end{equation}
with
\begin{equation}
    \lambda
    =
    \frac{m\omega_0}{\kappa}
    -
    \frac{s}{R_+}.
    \label{eq:usual_lambda}
\end{equation}
The first term comes from the oscillator coupling after the near-horizon rescaling, while the second term is the angular spinorial contribution.

This form follows from the decomposition of the separated Dirac equation, where the compact angular direction contributes through the eigenvalue $s/R_+$ and the nonminimal oscillator interaction provides the effective radial shift.

In the static one-scale notation, $\kappa=\alpha/\ell$ and $R_+=\alpha\ell$, Eq.~\eqref{eq:usual_lambda} becomes
\begin{equation}
    \lambda
    =
    \frac{m\omega_0\ell}{\alpha}
    -
    \frac{s}{\alpha\ell}.
    \label{eq:usual_lambda_alpha_limit}
\end{equation}

For the rotating BTZ geometry, $\kappa$ and $R_+$ represent independent near-horizon scales. Therefore, the coupling $\lambda$ should be kept in the form of Eq.~\eqref{eq:usual_lambda} unless the static limit is explicitly imposed.

Generalized Dirac oscillators replace the linear oscillator profile by a radial function and admit Cornell and singular choices in curved backgrounds \cite{deng2018}; Cornell-type generalized couplings have also been considered in the near-horizon BTZ two-body problem \cite{guvendi2024pair}. For the one-body system considered here, an effective covariant interaction adapted to the stationary congruence can be written by introducing the outward unit vector
\begin{equation}
    n^\mu
    =
    \frac{a^\mu}{\mathfrak{a}}
\end{equation}
and the local Clifford element $\gamma(u)=u_\mu\gamma^\mu$. We consider the effective nonminimal replacement
\begin{equation}
    \nabla_\mu
    \longrightarrow
    \nabla_\mu
    -
    \mathcal{W}(\rho)\,
    \gamma(u)\,n_\mu .
    \label{eq:covariant_generalized_oscillator}
\end{equation}
Because $u^\mu n_\mu=0$, the corresponding Dirac equation may be written as
\begin{equation}
    \left[
        i\gamma^\mu\nabla_\mu
        -m
        +
        \mathcal{W}(\rho)
        \sigma^{\mu\nu}u_\mu n_\nu
    \right]\Psi
    =
    0,
    \qquad
    \sigma^{\mu\nu}
    =
    \frac{i}{2}
    [\gamma^\mu,\gamma^\nu].
    \label{eq:covariant_generalized_dirac}
\end{equation}
In the corotating orthonormal frame of Eq.~\eqref{eq:dreibein}, with $u=e_0$ and $n=e_1$, the gamma-matrix conventions \eqref{eq:gamma_representation} give
\begin{equation}
    \sigma^{\mu\nu}u_\mu n_\nu
    =
    \gamma^2.
    \label{eq:covariant_coupling_gamma2}
\end{equation}
The interaction therefore enters precisely in the matrix channel already occupied by the angular term, and after separation
\begin{equation}
    \mathcal{L}(\rho)
    =
    \mathcal{W}(\rho)
    -
    \frac{s}{R_+}.
    \label{eq:L_from_covariant_W}
\end{equation}
For a constant $\mathcal{W}=m\omega_0/\kappa$, Eq.~\eqref{eq:L_from_covariant_W} reproduces Eq.~\eqref{eq:usual_lambda}. For real $\mathcal{W}$, the interaction matrix is Dirac-Hermitian, so the standard $U(1)$ Dirac current remains conserved.

The geometric profiles in Eq.~\eqref{eq:geometric_profiles} motivate the two singular interactions. Retaining their leading near-horizon pieces and absorbing regular constant contributions into the oscillator-like term yields precisely the family introduced in Eq.~\eqref{eq:intro_lambda_iso}. These local models are therefore the leading singular near-horizon truncations of the effective interaction \eqref{eq:covariant_generalized_dirac}; they leave the BTZ geometry and tetrad unchanged. We use the subscript $p$ consistently for model-dependent quantities, including the response $\mathcal{Z}_p$ and its intersection parameter $\gamma_p$.

The pole order also fixes the local analytical class. For $p=1$, the coefficient matrix of the first-order system has a simple pole at $\rho=0$, so the horizon is a regular singular point of a Fuchsian system. For $p>1$, the leading singular matrix is
\begin{equation}
    \frac{1}{\rho^p}
    \begin{pmatrix}
        -\beta_p & 0\\
        0 & \beta_p
    \end{pmatrix},
    \label{eq:general_p_leading_matrix}
\end{equation}
which gives an irregular singular point of Poincar\'e rank $p-1$ and the leading essential factors
\begin{equation}
    \exp\!\left[
        \pm
        \frac{\beta_p}{
            (p-1)\rho^{p-1}
        }
    \right].
    \label{eq:general_p_essential_factor}
\end{equation}
Thus $p=1$ is the Fuchsian threshold, while $p=2$ is the first irregular member of the singular family. Regular corrections in the full geometric profiles do not change this local classification.

\section{\texorpdfstring{Inverse-radial model ($p=1$)}{Inverse-radial model (p=1)}}
\label{sec:p1_isotonic}

\subsection{Isotonic motivation and inverse-radial coupling}
\label{subsec:p1_coupling_motivation}

The generic radial system derived in Sec.~\ref{sec:btz_dirac_system} shows that the near-horizon Dirac oscillator is determined by the effective coupling $\mathcal{L}_p(\rho)$ together with the near-horizon geometric parameters. In the reference oscillator problem this coupling is constant. We first retain the leading singular part of the geometric profile $F_1=\mathfrak{a}$, which produces the $p=1$ deformation on the half-line $\rho>0$. The resulting first-order system is Fuchsian.

The motivation comes from the isotonic oscillator, whose characteristic radial structure is associated with a superpotential containing both a linear term and an inverse-coordinate term,
\begin{equation}
    W(\rho)
    =
    c_{\rm lin}\rho+\frac{c_{\rm inv}}{\rho},
    \qquad
    \rho>0 .
    \label{eq:isotonic_superpotential_p1_section}
\end{equation}
In the near-horizon BTZ problem, the tetrad structure and spin connection determine how a nonminimal interaction enters the separated coupling. Equation~\eqref{eq:isotonic_superpotential_p1_section} therefore motivates a class of singular half-line interactions rather than a direct identification of $(c_{\rm lin},c_{\rm inv})$ with the radial parameters below. This distinction separates the present construction from flat-space Dirac-isotonic models \cite{agboola2012,ghosh2025}; we consequently use the term \emph{isotonic-inspired}. The isotonic half-line motivation and the stationary geometric profile $F_1\sim1/\rho$ therefore select the same leading inverse-radial structure.

The inverse-radial coupling is defined as
\begin{equation}
    \mathcal{L}_1(\rho)
    =
    \lambda_0+\frac{\beta_1}{\rho},
    \label{eq:lambda_iso_definition}
\end{equation}
where
\begin{equation}
    \lambda_0
    =
    \lambda_{\rm osc}
    -
    \frac{s}{R_+}.
    \label{eq:lambda0_definition}
\end{equation}
Here $\lambda_{\rm osc}$ denotes the regular constant oscillator-like contribution retained after the near-horizon reduction, and $\beta_1$ controls the inverse-radial deformation. For the reference Dirac oscillator, $\lambda_{\rm osc}=m\omega_0/\kappa$, consistently with Eq.~\eqref{eq:usual_lambda}. The constant-coupling radial system is recovered when $\beta_1\to0$.

The deformation in Eq.~\eqref{eq:lambda_iso_definition} should be understood as an isotonic-inspired Dirac coupling rather than the isotonic potential itself, since the inverse-radial term enters the first-order spinorial system and the associated componentwise scalar singular structure is exposed by the decoupling procedure.

Equation~\eqref{eq:lambda_iso_definition} is the leading singular near-horizon truncation of the effective interaction described in Sec.~\ref{subsec:generic_radial_coupling}; it remains an external interaction on the fixed BTZ background. A second coupling proportional to $\rho^{-2}$ is treated in Sec.~\ref{sec:p2_isotonic}. The two deformations have different parameter dimensions, pole orders, and horizon behavior; their comparison is therefore formulated in terms of the local coupling data at the matching radius.

Substituting Eq.~\eqref{eq:lambda_iso_definition} into the generic radial system gives
\begin{equation}
    \left[
        \epsilon(\rho)-m
    \right]\psi_1
    +
    \left[
        \lambda_0+\frac{\beta_1}{\rho}
        -
        \frac{d}{d\rho}
        -
        \frac{1}{2\rho}
    \right]\psi_2
    =
    0,
    \label{eq:p1_system_1}
\end{equation}
\begin{equation}
    \left[
        \epsilon(\rho)+m
    \right]\psi_2
    +
    \left[
        \lambda_0+\frac{\beta_1}{\rho}
        +
        \frac{d}{d\rho}
        +
        \frac{1}{2\rho}
    \right]\psi_1
    =
    0,
    \label{eq:p1_system_2}
\end{equation}
Equations~\eqref{eq:p1_system_1} and \eqref{eq:p1_system_2} define the inverse-radial isotonic-inspired deformation studied below.

\subsection{Decoupling of the coupled radial system}
\label{subsec:p1_decoupling}

The inverse-radial model is intrinsically a coupled first-order spinorial problem. A second-order equation can be obtained by eliminating one component; however, this procedure is valid only on intervals where the coefficient $\epsilon(\rho)+m$ does not vanish. The radial dependence of all coefficients must be retained throughout the calculation.

From Eq.~\eqref{eq:p1_system_2}, the second spinor component is
\begin{equation}
    \psi_2(\rho)
    =
    -
    \frac{
        \psi_1'(\rho)
        +
        \left[
            \mathcal{L}_1(\rho)
            +
            \frac{1}{2\rho}
        \right]\psi_1(\rho)
    }
    {\epsilon(\rho)+m}.
    \label{eq:psi2_from_psi1_p1}
\end{equation}
The denominator is
\begin{equation}
    \epsilon(\rho)+m
    =
    \frac{\Omega}{\rho}+m
    =
    \frac{\Omega+m\rho}{\rho}.
    \label{eq:epsilon_plus_m_p1}
\end{equation}
It is therefore a radial function. Its derivative contributes when Eq.~\eqref{eq:psi2_from_psi1_p1} is substituted into Eq.~\eqref{eq:p1_system_1}.

Let
\begin{equation}
    y(\rho)
    \equiv
    \psi_1(\rho).
    \label{eq:y_definition_p1}
\end{equation}
After substitution and simplification, the first component satisfies
\begin{equation}
    y''(\rho)
    +
    P(\rho)y'(\rho)
    +
    Q(\rho)y(\rho)
    =
    0,
    \label{eq:p1_second_order}
\end{equation}
with
\begin{equation}
    P(\rho)
    =
    \frac{2}{\rho}
    -
    \frac{m}{\Omega+m\rho},
    \label{eq:P_p1}
\end{equation}
and
\begin{align}
    Q(\rho)
    =
    \frac{1}
    {4\rho^2(\Omega+m\rho)}
    \Big\{
    &
    4\Omega^3
    +
    4m\Omega^2\rho
    -
    m\rho
    \left[
        (1+2\beta_1)^2
        +
        8\beta_1\lambda_0\rho
        +
        4(\lambda_0^2+m^2)\rho^2
    \right]
    \nonumber\\
    &
    -
    \Omega
    \left[
        -1+4\beta_1^{\,2}
        +
        4(-1+2\beta_1)\lambda_0\rho
        +
        4(\lambda_0^2+m^2)\rho^2
    \right]
    \Big\}.
    \label{eq:Q_p1}
\end{align}
Equation~\eqref{eq:p1_second_order} is the decoupled radial equation associated with the first spinor component.

The second component is not an independent solution of a separate scalar problem. Once $y(\rho)=\psi_1(\rho)$ is known on a domain where $\epsilon(\rho)+m\neq0$, $\psi_2(\rho)$ is fixed by Eq.~\eqref{eq:psi2_from_psi1_p1}. The physical radial mode is therefore the spinor pair reconstructed from the original coupled system.

\subsection{Singularity structure}
\label{subsec:p1_singularity_structure}

The singular structure of Eq.~\eqref{eq:p1_second_order} follows from the rational coefficients $P(\rho)$ and $Q(\rho)$. For generic $m\neq0$ and $\Omega\neq0$, the finite singular points are
\begin{equation}
    \rho=0,
    \qquad
    \Omega+m\rho=0.
    \label{eq:finite_singular_points_p1}
\end{equation}
The first point is the black-hole horizon in the near-horizon coordinate system. The second point is
\begin{equation}
    \rho_c
    =
    -\frac{\Omega}{m}.
    \label{eq:rho_c_p1}
\end{equation}
It is not a curvature singularity of the BTZ background. It is generated by the spinorial decoupling through the factor $\epsilon(\rho)+m$ in Eq.~\eqref{eq:epsilon_plus_m_p1}. For complex $\Omega$, this point should be understood as a singular point of the complexified radial equation and need not lie on the real integration interval.

The horizon $\rho=0$ is a regular singular point. To determine the local behavior, we set
\begin{equation}
    y(\rho)
    \sim
    \rho^q,
    \qquad
    \rho\to0 .
    \label{eq:near_horizon_power_law_p1}
\end{equation}
For $\Omega\neq0$, keeping the leading inverse powers of $\rho$ in Eq.~\eqref{eq:p1_second_order} gives the indicial equation
\begin{equation}
    \left(
        q+\frac{1}{2}
    \right)^2
    =
    \beta_1^{\,2}-\Omega^2.
    \label{eq:indicial_equation_p1}
\end{equation}
It is convenient to define
\begin{equation}
    \sigma
    \equiv
    \sqrt{\beta_1^{\,2}-\Omega^2},
    \qquad
    q_\pm
    =
    -\frac{1}{2}
    \pm
    \sigma .
    \label{eq:qpm_p1}
\end{equation}
For complex $\Omega$, the square-root branch in $\sigma$ is chosen so that $\operatorname{Re}\sigma\geq0$. When $\operatorname{Re}\sigma>0$, $q_+$ is the less-singular exponent; when $\operatorname{Re}\sigma=0$, the two exponents have equal real parts and $q_+$ denotes the continuation of the chosen branch. This convention is used below.

The spectral point $\Omega=0$ is exceptional for the componentwise scalar reduction because $\epsilon(\rho)+m$ no longer has the same leading near-horizon behavior used above. For $m\neq0$, direct substitution into the exact $\Omega=0$ scalar equation gives
\begin{equation}
    q^2-\left(\beta_1+\frac12\right)^2=0,
    \label{eq:p1_omega_zero_indicial}
\end{equation}
so the $\psi_1$ exponents are $q=\beta_1+1/2$ and $q=-\beta_1-1/2$. This exceptional componentwise result should not be obtained by naively setting $\Omega=0$ in Eq.~\eqref{eq:qpm_p1}; the complete first-order spinor system remains the primary description at this point.

For generic $m\neq0$ and $\Omega\neq0$, the second finite singular point, $\rho=\rho_c$, is also regular. Around this point, $P(\rho)$ has at most a simple pole and $Q(\rho)$ has at most a simple pole. Hence
\begin{equation}
    (\rho-\rho_c)P(\rho)
    \quad \text{and} \quad
    (\rho-\rho_c)^2Q(\rho)
\end{equation}
remain finite. This point is a feature of the decoupled spinorial equation and is absent from a reduction in which the denominator $\epsilon(\rho)+m$ is treated as constant.

The point at infinity is irregular. For large $\rho$, the coefficients behave as
\begin{equation}
    P(\rho)
    =
    \frac{1}{\rho}
    +
    O\left(
        \frac{1}{\rho^2}
    \right),
    \label{eq:P_asymptotic_p1}
\end{equation}
and
\begin{equation}
    Q(\rho)
    =
    -(\lambda_0^2+m^2)
    +
    O\left(
        \frac{1}{\rho}
    \right).
    \label{eq:Q_asymptotic_p1}
\end{equation}
The asymptotic equation therefore contains a nonzero constant radial scale and admits exponential behavior at infinity.

For generic $m\neq0$ and $\Omega\neq0$, the scalar equation for $\psi_1$ has two finite regular singular points and one irregular point at infinity and therefore belongs to the confluent-Heun class. The complete first-order system remains regular at $\rho_c$: the pole originates only from the division by $\epsilon(\rho)+m$ in Eq.~\eqref{eq:psi2_from_psi1_p1}. Therefore, $\rho_c$ should be interpreted as a component-dependent singularity of the decoupled scalar equation rather than as a singularity of the original first-order Dirac system.

\subsection{Constant-matrix form and Whittaker reduction}
\label{subsec:whittaker_reduction}

For the inverse-radial coupling, the complete spinor system takes the constant-matrix form
\begin{equation}
    \frac{d\Psi_{\rm rad}}{d\rho}
    =
    \left(
        A_1+\frac{A_0}{\rho}
    \right)\Psi_{\rm rad},
    \label{eq:constant_matrix_system}
\end{equation}
where
\begin{equation}
    A_0
    =
    \begin{pmatrix}
        -\beta_1-\frac12 & -\Omega\\
        \Omega & \beta_1-\frac12
    \end{pmatrix},
    \qquad
    A_1
    =
    \begin{pmatrix}
        -\lambda_0 & -m\\
        -m & \lambda_0
    \end{pmatrix}.
    \label{eq:A0_A1_matrices}
\end{equation}
Define the radial scale
\begin{equation}
    \Xi
    =
    \sqrt{\lambda_0^2+m^2},
    \qquad
    T=
    \begin{pmatrix}
        1 & 1\\[2mm]
        -\dfrac{\lambda_0+\Xi}{m}
        &
        \dfrac{\Xi-\lambda_0}{m}
    \end{pmatrix},
    \qquad
    \Phi=T^{-1}\Psi_{\rm rad}
    =
    \begin{pmatrix}u\\v\end{pmatrix},
    \label{eq:whittaker_transformation}
\end{equation}
for $m\neq0$. Direct multiplication gives
\begin{equation}
    T^{-1}A_1T
    =
    \begin{pmatrix}
        \Xi&0\\
        0&-\Xi
    \end{pmatrix}.
    \label{eq:A1_diagonal}
\end{equation}
Writing
\begin{equation}
    B=T^{-1}A_0T
    =
    \begin{pmatrix}
        B_{11}&B_{12}\\
        B_{21}&B_{22}
    \end{pmatrix},
    \label{eq:B_matrix}
\end{equation}
the matrix elements required below are
\begin{align}
    B_{11}&=-\frac12+\frac{\lambda_0\beta_1}{\Xi},
    &
    B_{22}&=-\frac12-\frac{\lambda_0\beta_1}{\Xi},
    \label{eq:B_diagonal_entries}\\
    B_{12}&=
    -\frac{(\Xi-\lambda_0)(\Omega\Xi+\beta_1m)}
    {\Xi m},
    &
    B_{21}&=
    \frac{(\Xi+\lambda_0)(\Omega\Xi-\beta_1m)}
    {\Xi m},
    \nonumber\\
    &&
    \det B&=
    \Omega^2-\beta_1^{\,2}+\frac14 .
    \label{eq:B_offdiag_determinant}
\end{align}
The transformed system is
\begin{equation}
    u'
    =
    \left(
        \Xi+\frac{B_{11}}{\rho}
    \right)u
    +
    \frac{B_{12}}{\rho}v,
    \qquad
    v'
    =
    \frac{B_{21}}{\rho}u
    +
    \left(
        -\Xi+\frac{B_{22}}{\rho}
    \right)v.
    \label{eq:transformed_first_order_system}
\end{equation}
For generic $B_{12}\neq0$, elimination of $v$ gives
\begin{equation}
    u''
    +
    \frac{2}{\rho}u'
    -
    \left[
        \Xi^2
        +
        \frac{\Xi(B_{11}-B_{22}+1)}{\rho}
        -
        \frac{\det B}{\rho^2}
    \right]u
    =
    0.
    \label{eq:u_whittaker_pre}
\end{equation}
With
\begin{equation}
    u(\rho)=\frac{w(z)}{\rho},
    \qquad
    z=2\Xi\rho,
    \qquad
    \kappa_{\rm W}
    =
    -\frac12-\frac{\lambda_0\beta_1}{\Xi},
    \label{eq:whittaker_variables}
\end{equation}
Eq.~\eqref{eq:u_whittaker_pre} becomes
\begin{equation}
    \frac{d^2w}{dz^2}
    +
    \left[
        -\frac14
        +
        \frac{\kappa_{\rm W}}{z}
        +
        \frac{\frac14-\sigma^2}{z^2}
    \right]w
    =
    0.
    \label{eq:whittaker_equation}
\end{equation}
This is the Whittaker equation. When $2\sigma\notin\mathbb Z$, a convenient local basis is
\begin{equation}
    u_\pm(\rho)
    =
    \frac{1}{\rho}
    M_{\kappa_{\rm W},\,\pm\sigma}(2\Xi\rho),
    \label{eq:u_whittaker_solutions}
\end{equation}
where $M_{\kappa,\mu}$ is the Whittaker $M$ function.
For degenerate values satisfying $2\sigma\in\mathbb Z$, the second independent solution is obtained by the standard limiting procedure or by using the Whittaker $W$ function.

The second transformed component and the physical spinor follow from
\begin{equation}
    v_\pm(\rho)
    =
    \frac{\rho}{B_{12}}
    \left[
        u_\pm'(\rho)
        -
        \left(
            \Xi+\frac{B_{11}}{\rho}
        \right)u_\pm(\rho)
    \right],
    \qquad
    \Psi_{{\rm rad},\pm}=T
    \begin{pmatrix}u_\pm\\v_\pm\end{pmatrix}.
    \label{eq:whittaker_spinor_reconstruction}
\end{equation}
Since $M_{\kappa,\sigma}(z)\sim z^{\sigma+1/2}$ as $z\to0$, the plus branch has the near-horizon scaling
\begin{equation}
    \Psi_{{\rm rad},+}
    \propto
    \rho^{-1/2+\sigma},
\end{equation}
up to a constant spinor prefactor determined by $T$, and reproduces the $q_+$ branch of Eq.~\eqref{eq:qpm_p1}.

If $B_{12}=0$, the reconstruction formula above is not applicable. The system becomes triangular in the transformed basis and the second component must be obtained directly from the first-order equations, or equivalently by eliminating the opposite component.

The cases $m=0$ or $\Xi=0$ require a separate, simpler diagonalization and are not covered by the parametrization \eqref{eq:whittaker_transformation}.

The standard constant-coupling near-horizon Dirac oscillator is recovered continuously by setting $\beta_1=0$. In this limit, Eq.~\eqref{eq:qpm_p1} gives $\sigma=\sqrt{-\Omega^2}$ and Eq.~\eqref{eq:whittaker_variables} gives $\kappa_{\rm W}=-1/2$. The generic system therefore remains within the same Whittaker class in the rotated spinor basis, with the degenerate cases treated directly from the first-order equations as described above. Further details of the connection with the reference oscillator are collected in Sec.~\ref{subsec:usual_oscillator_limit}.

\section{\texorpdfstring{$\rho^{-2}$ model ($p=2$)}{Inverse-square model (p=2)}}
\label{sec:p2_isotonic}

\subsection{\texorpdfstring{Geometric motivation and definition of the $p=2$ coupling}{Geometric motivation and definition of the p=2 coupling}}
\label{subsec:p2_origin}

The $p=1$ model of Sec.~\ref{sec:p1_isotonic} retains the leading profile $F_1\sim\rho^{-1}$ and remains Fuchsian. The second geometric profile in Eq.~\eqref{eq:geometric_profiles} satisfies $F_2=-\mathfrak{a}'\sim\rho^{-2}$ and therefore motivates the next member of the singular family. This step is qualitatively different from adding a small correction to $p=1$: a pole of order two is the first irregular case of the first-order radial system.

The isotonic superpotential introduced in Eq.~\eqref{eq:isotonic_superpotential_p1_section} provides the same half-line motivation for the $p=2$ model.
The $p=2$ member is not the standard isotonic superpotential itself. Its $\rho^{-2}$ term is instead the leading singular contribution associated with the second geometric profile, $F_2=-\mathfrak{a}'=(2\pi T_{\rm loc})^2$. We therefore define the local near-horizon model by
\begin{equation}
    \mathcal{L}_2(\rho)
    =
    \lambda_0+\frac{\beta_2}{\rho^2}.
    \label{eq:lambda_p2_definition}
\end{equation}
Here $\lambda_0$ is the same constant oscillator-like contribution used for $p=1$, while $\beta_2$ controls the inverse-square radial deformation.

Substituting Eq.~\eqref{eq:lambda_p2_definition} into the generic radial system gives
\begin{equation}
    \left[
        \epsilon(\rho)-m
    \right]\psi_1
    +
    \left[
        \lambda_0+\frac{\beta_2}{\rho^2}
        -
        \frac{d}{d\rho}
        -
        \frac{1}{2\rho}
    \right]\psi_2
    =
    0,
    \label{eq:p2_system_1}
\end{equation}
\begin{equation}
    \left[
        \epsilon(\rho)+m
    \right]\psi_2
    +
    \left[
        \lambda_0+\frac{\beta_2}{\rho^2}
        +
        \frac{d}{d\rho}
        +
        \frac{1}{2\rho}
    \right]\psi_1
    =
    0,
    \label{eq:p2_system_2}
\end{equation}

The $p=2$ model is not a notational variant of $p=1$. The term $\beta_2/\rho^2$ is more singular than both the frequency contribution $\epsilon(\rho)$
and the spin-connection contribution proportional to $1/(2\rho)$
near the horizon.
It therefore changes the dominant local structure and the classification of the singular point from Fuchsian to irregular. Equation~\eqref{eq:lambda_p2_definition} is the leading singular truncation of the effective interaction described in Sec.~\ref{subsec:generic_radial_coupling}; the regular terms omitted from $F_2$ do not modify the Poincar\'e rank or the essential near-horizon behavior.

\subsection{Essential near-horizon factor}
\label{subsec:p2_essential_factor}

The dominant singular terms of Eqs.~\eqref{eq:p2_system_1} and
\eqref{eq:p2_system_2} show that the physical spinor components contain an essential near-horizon factor. To isolate this behavior, we write
\begin{equation}
    \psi_i(\rho)
    =
    e^{\beta_2/\rho}\rho^{-1/2}u_i(\rho),
    \qquad
    i=1,2.
    \label{eq:p2_factorization}
\end{equation}
The logarithmic derivative of the common prefactor is
\begin{equation}
    \frac{d}{d\rho}
    \ln
    \left(
        e^{\beta_2/\rho}\rho^{-1/2}
    \right)
    =
    -\frac{\beta_2}{\rho^2}
    -
    \frac{1}{2\rho}.
    \label{eq:p2_prefactor_derivative}
\end{equation}
Thus, the essential exponential accounts for the leading $\rho^{-2}$ behavior, while the factor $\rho^{-1/2}$ compensates the spin-connection contribution proportional to $1/(2\rho)$.

Substitution of Eq.~\eqref{eq:p2_factorization} into the first-order system gives the auxiliary equations
\begin{equation}
    u_1'(\rho)
    =
    -
    \left[
        \epsilon(\rho)+m
    \right]u_2(\rho)
    -
    \lambda_0 u_1(\rho),
    \label{eq:p2_u_system_1}
\end{equation}
\begin{equation}
    u_2'(\rho)
    =
    \left[
        \epsilon(\rho)-m
    \right]u_1(\rho)
    +
    \left[
        \lambda_0+\frac{2\beta_2}{\rho^2}
    \right]u_2(\rho).
    \label{eq:p2_u_system_2}
\end{equation}

The above transformation removes the explicit essential exponential contribution associated with the chosen branch, but it does not eliminate every $\rho^{-2}$ term from the auxiliary system. In particular, Eq.~\eqref{eq:p2_u_system_2} retains the coefficient $2\beta_2/\rho^2$. The role of the factorization is therefore to extract the dominant essential behavior from the physical spinor and leave an auxiliary system suitable for a local formal asymptotic expansion.

The factorization \eqref{eq:p2_factorization} describes only one of the two essential branches. The complementary branch is obtained from
\begin{equation}
    \psi_i(\rho)
    =
    e^{-\beta_2/\rho}\rho^{-1/2}\widetilde u_i(\rho),
    \qquad i=1,2,
    \label{eq:p2_second_factorization}
\end{equation}
which gives
\begin{equation}
    \widetilde u_1'
    =
    -
    \left(\epsilon+m\right)\widetilde u_2
    -
    \left(
        \lambda_0
        +
        \frac{2\beta_2}{\rho^2}
    \right)\widetilde u_1,
    \qquad
    \widetilde u_2'
    =
    \left(\epsilon-m\right)\widetilde u_1
    +
    \lambda_0\widetilde u_2 .
    \label{eq:p2_second_auxiliary_system}
\end{equation}

For the first branch, $u_1$ is dominant and $u_2=O(\rho)$; for the complementary branch, $\widetilde u_2$ is dominant and $\widetilde u_1=O(\rho)$. The two essential branches are related by the sign of the exponential factor. For a fixed sign of $\beta_2$, the locally bounded branch is the one proportional to $e^{-|\beta_2|/\rho}\rho^{-1/2}$. The dominant component assignment depends on the chosen branch and the sign of $\beta_2$; consequently, it should be interpreted as an asymptotic statement rather than an invariant spinor property.

With the normalization $\widetilde u_{2,0}=1$, the complementary branch begins as
\begin{equation}
    \widetilde u_2(\rho)=1+O(\rho),
    \qquad
    \widetilde u_1(\rho)
    =
    -\frac{\Omega}{2\beta_2}\rho
    +
    O(\rho^2).
    \label{eq:p2_second_branch_leading}
\end{equation}

The essential factors $e^{\pm\beta_2/\rho}\rho^{-1/2}$ are defined with respect to the chosen corotating near-horizon frame used in the separated radial problem. Their exponential suppression or growth as $\rho\to0^+$ should therefore not, by itself, be interpreted as a complete regularity criterion for a freely falling observer crossing the horizon. The effective interaction in Sec.~\ref{subsec:generic_radial_coupling} is adapted to the stationary congruence, whose proper acceleration diverges as the horizon is approached; an infalling regularity criterion instead requires a horizon-regular coordinate system and spin frame. The locally bounded essential branch is therefore selected to define finite-radius response data, rather than to impose a global quasinormal-mode or infalling boundary condition.

\subsection{Formal asymptotic expansion}
\label{subsec:p2_regularized_series}

The auxiliary system \eqref{eq:p2_u_system_1}--\eqref{eq:p2_u_system_2} can be converted into a second-order equation for
\begin{equation}
    y(\rho)
    \equiv
    u_1(\rho).
    \label{eq:p2_y_def}
\end{equation}
Eliminating $u_2$ from Eq.~\eqref{eq:p2_u_system_1}, one obtains
\begin{equation}
    y''(\rho)
    +
    P_u(\rho)y'(\rho)
    +
    Q_u(\rho)y(\rho)
    =
    0,
    \label{eq:p2_second_order_u}
\end{equation}
where
\begin{equation}
    P_u(\rho)
    =
    \frac{1}{\rho}
    -
    \frac{m}{\Omega+m\rho}
    -
    \frac{2\beta_2}{\rho^2},
    \label{eq:Pu_p2}
\end{equation}
and
\begin{equation}
    Q_u(\rho)
    =
    \frac{\Omega^2-2\beta_2\lambda_0}{\rho^2}
    +
    \frac{\lambda_0}{\rho}
    -
    \lambda_0^2
    -
    m^2
    -
    \frac{\lambda_0 m}{\Omega+m\rho}.
    \label{eq:Qu_p2}
\end{equation}
Although Eq.~\eqref{eq:p2_second_order_u} remains irregular at the horizon, the dominant essential factor has been isolated in Eq.~\eqref{eq:p2_factorization}. The auxiliary functions admit formal expansions
\begin{equation}
    u_1(\rho)
    =
    \sum_{n=0}^{\infty}a_n\rho^n,
    \qquad
    u_2(\rho)
    =
    \sum_{n=0}^{\infty}b_n\rho^n.
    \label{eq:p2_series_ansatz}
\end{equation}
Substitution of Eq.~\eqref{eq:p2_series_ansatz} into
Eqs.~\eqref{eq:p2_u_system_1} and \eqref{eq:p2_u_system_2} gives, from the lowest powers of $\rho$,
\begin{equation}
    b_0=0,
    \qquad
    b_1
    =
    -\frac{\Omega}{2\beta_2}a_0.
    \label{eq:p2_lowest_coefficients}
\end{equation}
The overall normalization is arbitrary. We choose
\begin{equation}
    a_0=1.
    \label{eq:p2_a0_normalization}
\end{equation}

For $k\geq0$, the recurrence relations are
\begin{equation}
    (k+1)a_{k+1}
    +
    \lambda_0 a_k
    +
    \Omega b_{k+1}
    +
    m b_k
    =
    0,
    \label{eq:p2_recurrence_a}
\end{equation}
and
\begin{equation}
    2\beta_2 b_{k+2}
    =
    (k+1)b_{k+1}
    -
    \Omega a_{k+1}
    +
    m a_k
    -
    \lambda_0 b_k .
    \label{eq:p2_recurrence_b}
\end{equation}
These relations determine a formal local expansion order by order for $\beta_2\neq0$. They do not, in general, define convergent Taylor series. Indeed, the late-order part of Eq.~\eqref{eq:p2_recurrence_b} contains
\begin{equation}
    b_{k+2}
    \sim
    \frac{k+1}{2\beta_2}b_{k+1},
    \qquad
    k\to\infty,
    \label{eq:p2_factorial_growth}
\end{equation}
which implies factorial coefficient growth for generic parameters. The expansions are therefore asymptotic (of zero radius of convergence in the generic case) and should be truncated near their least term. The case $\beta_2=0$ is singular with respect to this expansion and must be solved directly as the constant-coupling system.

The first essential branch consequently has the formal asymptotic form
\begin{equation}
    \psi_1(\rho)
    =
    e^{\beta_2/\rho}\rho^{-1/2}
    \sum_{n=0}^{\infty}a_n\rho^n,
    \label{eq:p2_psi1_series}
\end{equation}
\begin{equation}
    \psi_2(\rho)
    =
    e^{\beta_2/\rho}\rho^{-1/2}
    \sum_{n=0}^{\infty}b_n\rho^n,
    \label{eq:p2_psi2_series}
\end{equation}
with the coefficients generated by
Eqs.~\eqref{eq:p2_lowest_coefficients}--\eqref{eq:p2_recurrence_b}. The complementary recurrence follows by substituting independent formal series into Eq.~\eqref{eq:p2_second_auxiliary_system}; its dominant coefficient is $\widetilde u_{2,0}$ rather than $u_{1,0}$. This structure distinguishes the $p=2$ model from the inverse-radial problem, whose complete spinor is exactly reducible to Whittaker functions.

\section{Finite-radius spinor response}
\label{sec:finite_radius_response}

\subsection{Damped-frequency slice}
\label{subsec:purely_imaginary_sector}

To analyze the finite-radius response on a one-dimensional frequency slice, we set the corotating frequency to
\begin{equation}
    \Omega=-i\gamma,
    \qquad
    \gamma>0,
    \label{eq:purely_imaginary_Omega}
\end{equation}
The corotating frequency is therefore
\begin{equation}
    \omega=-i\kappa\gamma .
    \label{eq:physical_frequency_damped}
\end{equation}
For a rotating BTZ black hole, Eq.~\eqref{eq:frequency_frame_relation} gives
\begin{equation}
    \omega_{\rm BL}=s\Omega_H-i\kappa\gamma .
    \label{eq:BL_damped_slice}
\end{equation}
Thus, the slice \eqref{eq:purely_imaginary_Omega} is purely imaginary only in the corotating frame. It defines a one-dimensional section of the complex response function on the near-horizon interval specified in Eq.~\eqref{eq:near_horizon_domain}.

For the $p=1$ model, the less-singular near-horizon exponent becomes
\begin{equation}
    q_+^{(1)}
    =
    -\frac{1}{2}
    +
    \sqrt{\beta_1^{\,2}+\gamma^2},
    \label{eq:qplus_damped_p1}
\end{equation}
because $\Omega^2=-\gamma^2$. The corresponding projective ratio follows directly from the leading first-order system,
\begin{equation}
    \frac{\psi_2}{\psi_1}
    \longrightarrow
    -
    \frac{
        q_+^{(1)}+\beta_1+\frac{1}{2}
    }
    {\Omega},
    \qquad
    \rho\to0^+ .
    \label{eq:p1_initial_ratio_damped}
\end{equation}
For the $p=2$ model, the analogous local ratio is determined by the factorized auxiliary systems and the formal expansions of Sec.~\ref{subsec:p2_regularized_series}. Which essential branch is locally bounded depends on the sign of $\beta_2$.

\subsection{Finite-radius response and flux}
\label{subsec:mixed_spinorial_boundary_condition}

The natural output of each local radial problem is the complex spinor response defined in Eq.~\eqref{eq:spinor_response_definition}, with $p=1,2$. Here $\psi_{i,p}$ denotes the $i$th component associated with the coupling $\mathcal{L}_p$. The response is a local near-horizon matching datum; by itself it is neither a quasinormal-mode condition nor a global spectral condition. A global condition requires matching to an exterior solution together with admissible asymptotic boundary data or to a separately derived finite-radius wall condition. The calligraphic symbol $\mathcal{Z}_p$ is reserved throughout for this spinorial response; the ordinary symbol $Z(x)$ denotes only the auxiliary confluent-Heun function introduced in Appendix~\ref{sec:heun_reconstruction}.

To define a real one-parameter section of the response, we impose
\begin{equation}
    \operatorname{Re}\mathcal{Z}_p(\Omega;\rho_0)
    =
    \eta,
    \qquad
    \eta\in\mathbb{R}.
    \label{eq:response_projection}
\end{equation}
This fixes one real component of the local response. Its relation to radial flux follows from the Dirac current. In the gamma-matrix representation \eqref{eq:gamma_representation},
\begin{equation}
    j_p^\rho
    =
    \overline\Psi_p\gamma^\rho\Psi_p
    =
    -\Psi_p^\dagger\sigma^2\Psi_p
    =
    -2|\psi_{1,p}|^2
    \operatorname{Im}\mathcal{Z}_p .
    \label{eq:radial_dirac_current}
\end{equation}
Consequently, a fluxless wall requires $\operatorname{Im}\mathcal{Z}_p=0$ in addition to Eq.~\eqref{eq:response_projection}. The distinction between a projected response, a flux-compatible wall condition, and a complete asymptotically AdS spectral problem is essential when imposing Robin or other admissible boundary data \cite{bussola2017,dappiaggi2018,wang2018,wang2020,konewko2024}; general admissibility in asymptotically AdS settings is discussed in Ref.~\cite{deoliveira2022bc}, while an acoustic-BTZ Robin construction is given in Ref.~\cite{deoliveira2023robin}.

For the $p=2$ model, each essential factorization in Eqs.~\eqref{eq:p2_factorization} and \eqref{eq:p2_second_factorization} multiplies both spinor components by the same scalar prefactor, which cancels from the projective coordinate. For the branch of Eq.~\eqref{eq:p2_factorization}, one has at every finite $\rho_0>0$,
\begin{equation}
    \mathcal{Z}_2(\Omega;\rho_0)
    =
    \frac{u_2(\rho_0)}{u_1(\rho_0)} .
    \label{eq:p2_ratio_equivalence}
\end{equation}
The physical response still depends on which local branch is selected, but not on the common scalar factor used to represent that branch.

\subsection{Analytical large-damping behavior}
\label{subsec:comparison_damping_times}

Wherever $\psi_{1,p}\neq0$, the projective coordinate \eqref{eq:spinor_response_definition} obeys an exact Riccati equation. Using Eqs.~\eqref{eq:generic_radial_system_1} and \eqref{eq:generic_radial_system_2}, the spin-connection terms cancel and one obtains
\begin{equation}
    \mathcal{Z}_p'
    =
    \left(\epsilon-m\right)
    +2\mathcal{L}_p\mathcal{Z}_p
    +\left(\epsilon+m\right)\mathcal{Z}_p^2 .
    \label{eq:response_riccati}
\end{equation}
For real $m$ and $\mathcal{L}_p$, and for the local branch satisfying $\mathcal{Z}_p=-i+O(\gamma^{-1})$ as $\gamma\to\infty$, a formal expansion of Eq.~\eqref{eq:response_riccati} gives at the outer matching radius
\begin{align}
    \operatorname{Re}\mathcal{Z}_p(-i\gamma;\rho_0)
    ={}&
    -\frac{m\rho_0}{\gamma}
    +\frac{m\rho_0\left(1-2\rho_0\mathcal{L}_{p,0}\right)}{2\gamma^2}
    \nonumber\\
    &+
    \frac{m\rho_0}{4\gamma^3}
    \left[
        -1+4\rho_0\mathcal{L}_{p,0}
        +2\rho_0^2
        \left(
            m^2-\mathcal{L}_{p,0}^{\,2}
            +\mathcal{L}'_{p,0}
        \right)
    \right]
    +O(\gamma^{-4}),
    \label{eq:response_large_gamma_general}
\end{align}
where
\begin{equation}
    \mathcal{L}_{p,0}=\mathcal{L}_p(\rho_0),
    \qquad
    \mathcal{L}'_{p,0}=\left.\frac{d\mathcal{L}_p}{d\rho}\right|_{\rho_0} .
    \label{eq:response_outer_coupling_data}
\end{equation}
The leading term depends only on the Dirac mass and the matching radius. The coefficient at order $\gamma^{-2}$ probes the value of the interaction at $\rho_0$, while its radial gradient first enters at order $\gamma^{-3}$.

This hierarchy becomes especially transparent when the two interactions are matched at the outer radius,
\begin{equation}
    \mathcal{L}_{1,0}
    =
    \mathcal{L}_{2,0}
    \equiv
    \mathcal{L}_0 .
    \label{eq:response_matched_outer_couplings}
\end{equation}
Then the first difference between the two projected responses is
\begin{equation}
    \operatorname{Re}\mathcal{Z}_1
    -
    \operatorname{Re}\mathcal{Z}_2
    =
    \frac{m\rho_0^3}{2\gamma^3}
    \left(
        \mathcal{L}'_{1,0}-\mathcal{L}'_{2,0}
    \right)
    +O(\gamma^{-4}).
    \label{eq:response_matched_difference}
\end{equation}
Thus interactions with the same outer value remain distinguishable through their local radial slope.

For $\eta<0$, write $e=|\eta|$ and assume $m>0$. Define the response-expansion coefficients
\begin{align}
    \mathcal{A}_{\rm R}&=m\rho_0,
    \\
    \mathcal{B}_{\rm R}&=\frac{m\rho_0}{2}
    \left(1-2\rho_0\mathcal{L}_0\right),
    \\
    \mathcal{C}_{{\rm R},p}&=\frac{m\rho_0}{4}
    \left[
        -1+4\rho_0\mathcal{L}_0
        +2\rho_0^2
        \left(
            m^2-\mathcal{L}_0^2+\mathcal{L}'_{p,0}
        \right)
    \right].
    \label{eq:response_ABC_coefficients}
\end{align}
Solving the projected condition $\operatorname{Re}\mathcal{Z}_p=-e$ asymptotically gives
\begin{equation}
    \gamma_p
    =
    \frac{\mathcal{A}_{\rm R}}{e}
    -\frac{\mathcal{B}_{\rm R}}{\mathcal{A}_{\rm R}}
    -\frac{\mathcal{A}_{\rm R}\mathcal{C}_{{\rm R},p}
    +\mathcal{B}_{\rm R}^{\,2}}
    {\mathcal{A}_{\rm R}^{\,3}}\,e
    +O(e^2).
    \label{eq:response_root_asymptotics}
\end{equation}
Consequently, for the matched outer couplings \eqref{eq:response_matched_outer_couplings},
\begin{equation}
    \Delta\gamma
    \equiv
    \gamma_1-\gamma_2
    =
    \frac{\rho_0}{2m}
    \left(
        \mathcal{L}'_{2,0}-\mathcal{L}'_{1,0}
    \right)e
    +O(e^2),
    \label{eq:response_delta_gamma_general}
\end{equation}
and
\begin{equation}
    1-\frac{\gamma_2}{\gamma_1}
    =
    \frac{
        \mathcal{L}'_{2,0}-\mathcal{L}'_{1,0}
    }{2m^2}
    e^2
    +O(e^3).
    \label{eq:response_splitting_asymptotics}
\end{equation}

For the explicit power-law models considered here, the common constant term $\lambda_0$ makes the matching condition \eqref{eq:response_matched_outer_couplings} equivalent to
\begin{equation}
    \frac{\beta_1}{\rho_0}
    =
    \frac{\beta_2}{\rho_0^2},
    \qquad\Longrightarrow\qquad
    \beta_2
    =
    \beta_1\rho_0 .
    \label{eq:matched_beta_relation}
\end{equation}
Their outer gradients then satisfy
\begin{equation}
    \mathcal{L}'_{2,0}
    -
    \mathcal{L}'_{1,0}
    =
    -\frac{\beta_1}{\rho_0^2},
    \label{eq:matched_gradient_explicit}
\end{equation}
and Eqs.~\eqref{eq:response_delta_gamma_general} and \eqref{eq:response_splitting_asymptotics} reduce to
\begin{align}
    \Delta\gamma
    &=
    -\frac{\beta_1}{2m\rho_0}\,e
    +O(e^2),
    \label{eq:explicit_delta_gamma}
    \\
    1-\frac{\gamma_2}{\gamma_1}
    &=
    -\frac{\beta_1}{2m^2\rho_0^2}\,e^2
    +O(e^3).
    \label{eq:explicit_relative_splitting}
\end{align}
More generally, the singular part $\delta\mathcal{L}_p=\beta_p/\rho^p$ obeys
\begin{equation}
    \rho\,
    \frac{\delta\mathcal{L}_p'}{\delta\mathcal{L}_p}
    =
    -p .
    \label{eq:logarithmic_slope_p}
\end{equation}
Thus, after the interaction strengths are matched at the outer surface, the next local datum entering the response retains direct information about the pole order.

Equations~\eqref{eq:response_delta_gamma_general} and \eqref{eq:response_splitting_asymptotics} are purely local analytical statements. They show that, after matching the outer values of the interactions, the absolute splitting of the projected response intersections is linear in $|\eta|$, whereas the relative splitting is quadratic, with coefficients fixed by the difference of the radial gradients at the matching surface.

The functions $\mathcal{Z}_p(\Omega;\rho_0)$ are local outputs of the near-horizon problems. Only after matching either response to the complete BTZ exterior, together with an admissible horizon branch and asymptotic AdS data, does one obtain a global spectral condition.

\section{Conclusions}\label{sec:conclusions}

The two singular generalized-oscillator profiles studied here admit a common near-horizon interpretation. For the stationary congruence generated by the BTZ horizon Killing field, the proper acceleration has leading behavior $1/\rho$, while its radial gradient satisfies $-\mathfrak{a}'=\mathfrak{a}^2-\ell^{-2}=(2\pi T_{\rm loc})^2$ and begins as $1/\rho^2$. The effective generalized-oscillator interaction introduced above reproduces the same matrix channel as the separated radial coupling, so $\mathcal{L}_p=\lambda_0+\beta_p/\rho^p$ represents the leading singular near-horizon form on the fixed BTZ background.

The pole order determines the analytical class of the spinor equation. The $p=1$ model is Fuchsian and forms the regular-singular threshold of the family; its matrix-Coulomb system has an exact Whittaker spinor, while the original component basis gives an equivalent confluent-Heun representation with an apparent finite singularity. The $p=2$ model is the first irregular member and produces the essential branches $e^{\pm\beta_2/\rho}\rho^{-1/2}$ together with formal asymptotic series governed by factorial late-order growth. More generally, a leading $\rho^{-p}$ interaction with $p>1$ has Poincar\'e rank $p-1$ and essential behavior $\exp[\pm\beta_p/((p-1)\rho^{p-1})]$.

The finite-radius response provides a direct analytical comparison of the two local singularity classes. Its exact Riccati equation gives $\operatorname{Re}\mathcal{Z}_p=-m\rho_0/\gamma+O(\gamma^{-2})$ on the slice $\Omega=-i\gamma$: the interaction value first appears at order $\gamma^{-2}$ and its radial gradient at order $\gamma^{-3}$. When the two explicit power-law couplings are matched at $\rho_0$, one has $\beta_2=\beta_1\rho_0$ and $\mathcal{L}'_{2,0}-\mathcal{L}'_{1,0}=-\beta_1/\rho_0^2$. The resulting splitting is therefore a direct local discriminator of pole order even after the interaction strengths have been equalized at the matching surface.

The finite-radius ratios $\mathcal{Z}_p(\Omega;\rho_0)$ provide the natural interface between the near-horizon solutions and the exterior BTZ problem. This effective construction does not turn the locally bounded branch into an infalling boundary condition: a global fermionic spectrum still requires a horizon-regular prescription and matching to the complete exterior with admissible AdS boundary data. The exact and asymptotic local solutions obtained here supply analytical input for that construction.

\section*{Acknowledgements}

This study was financed in part by the Coordenação de Aperfeiçoamento de Pessoal de Nível Superior - Brasil (CAPES) - Finance Code 001.

\section*{Data availability}

No data were created or analyzed in this study.

\appendix

\section{Componentwise confluent-Heun representation}
\label{sec:heun_reconstruction}

\subsection{Transformation to the Heun variable}
\label{subsec:heun_variable_transformation}

Eliminating $\psi_2$ in the original spinor basis gives an equivalent confluent-Heun representation of the first component. This form is convenient for boundary data specified directly for $\psi_1$. Throughout this appendix, the ordinary symbol $Z(x)$ denotes the auxiliary Heun function and remains distinct from the spinorial response $\mathcal{Z}_p$. For $m\neq0$ and $\Omega\neq0$, the finite singularities are mapped to the canonical locations by introducing the dimensionless variable
\begin{equation}
    x
    =
    -\frac{m\rho}{\Omega}.
    \label{eq:heun_variable}
\end{equation}
This transformation sends
\begin{equation}
    \rho=0
    \longmapsto
    x=0,
    \qquad
    \rho=-\frac{\Omega}{m}
    \longmapsto
    x=1,
    \qquad
    \rho=\infty
    \longmapsto
    x=\infty .
    \label{eq:heun_singularity_map}
\end{equation}
The derivatives transform as
\begin{equation}
    \frac{d}{d\rho}
    =
    -\frac{m}{\Omega}
    \frac{d}{dx},
    \qquad
    \frac{d^2}{d\rho^2}
    =
    \frac{m^2}{\Omega^2}
    \frac{d^2}{dx^2}.
    \label{eq:rho_x_derivatives}
\end{equation}

After substituting Eq.~\eqref{eq:heun_variable} into the decoupled equation
\eqref{eq:p1_second_order}, the first spinor component satisfies
\begin{equation}
    \frac{d^2y}{dx^2}
    +
    \left(
        \frac{2}{x}
        -
        \frac{1}{x-1}
    \right)
    \frac{dy}{dx}
    +
    \left(
        D_{\rm H}
        +
        \frac{A_{\rm H}}{x^2}
        +
        \frac{B_{\rm H}}{x}
        +
        \frac{C_{\rm H}}{x-1}
    \right)y
    =
    0.
    \label{eq:y_equation_x}
\end{equation}
The coefficients are
\begin{equation}
    A_{\rm H}
    =
    \Omega^2-\beta_1^{\,2}+\frac{1}{4},
    \label{eq:A_heun_pre}
\end{equation}
\begin{equation}
    B_{\rm H}
    =
    \frac{
        4\Omega\beta_1\lambda_0
        -
        2\Omega\lambda_0
        +
        2\beta_1m
        +
        m
    }
    {2m},
    \label{eq:B_heun_pre}
\end{equation}
\begin{equation}
    C_{\rm H}
    =
    \frac{
        2\Omega\lambda_0
        -
        2\beta_1m
        -
        m
    }
    {2m},
    \label{eq:C_heun_pre}
\end{equation}
and
\begin{equation}
    D_{\rm H}
    =
    -
    \frac{\Omega^2}{m^2}
    \left(
        \lambda_0^2+m^2
    \right).
    \label{eq:D_heun_pre}
\end{equation}
Equation~\eqref{eq:y_equation_x} contains a second-order pole at $x=0$ and a nonzero constant term. These terms are removed by extracting the dominant near-horizon power and the asymptotic exponential factor from the first spinor component.

\subsection{Heun parameters}
\label{subsec:heun_parameters}

Using the less-singular exponent $q_+$ from Eq.~\eqref{eq:qpm_p1} and the radial scale $\Xi$ from Eq.~\eqref{eq:whittaker_transformation}, we write
\begin{equation}
    y(x)
    =
    x^{q_+}e^{\tau_{\rm H}x}Z(x),
    \qquad
    \tau_{\rm H}
    =
    \frac{\Omega}{m}\Xi .
    \label{eq:heun_factorization}
\end{equation}
The power $q_+$ cancels the coefficient of $x^{-2}$ in the transformed equation, while $\tau_{\rm H}$ cancels the constant asymptotic term. The opposite sign, $\tau_{\rm H}\to-\tau_{\rm H}$, gives an equivalent confluent-Heun parametrization.

With Eq.~\eqref{eq:heun_factorization}, the function $Z(x)$ satisfies
\begin{equation}
    Z''(x)
    +
    \left[
        \alpha_{\rm H}
        +
        \frac{\beta_{\rm H}+1}{x}
        +
        \frac{\gamma_{\rm H}+1}{x-1}
    \right]Z'(x)
    +
    \left[
        \frac{q_{\rm H}}{x}
        +
        \frac{p_{\rm H}}{x-1}
    \right]Z(x)
    =
    0.
    \label{eq:confluent_heun_equation}
\end{equation}
This is the confluent-Heun equation in a form adapted to the present radial problem. The parameters are
\begin{equation}
    \alpha_{\rm H}
    =
    2\tau_{\rm H},
    \label{eq:alphaH}
\end{equation}
\begin{equation}
    \beta_{\rm H}
    =
    2\sigma,
    \label{eq:betaH}
\end{equation}
\begin{equation}
    \gamma_{\rm H}
    =
    -2,
    \label{eq:gammaH}
\end{equation}
\begin{equation}
    q_{\rm H}
    =
    \frac{
        -\lambda_0\Omega
        +
        m\sigma
        +
        \beta_1\left(m+2\lambda_0\Omega\right)
        +
        \Omega\Xi\left(1+2\sigma\right)
    }
    {m},
    \label{eq:qH}
\end{equation}
and
\begin{equation}
    p_{\rm H}
    =
    \frac{
        \left(\lambda_0-\Xi\right)\Omega
        -
        m\left(\beta_1+\sigma\right)
    }
    {m}.
    \label{eq:pH}
\end{equation}
The symbol $\beta_{\rm H}$ denotes a confluent-Heun parameter and should not be confused with the inverse-radial deformation parameter $\beta_1$.

Since $y=\psi_1$, Eq.~\eqref{eq:heun_factorization} directly represents the first spinor component, with $x$ defined in Eq.~\eqref{eq:heun_variable}. Here $Z(x)$ is a local solution of Eq.~\eqref{eq:confluent_heun_equation}. The branch relevant to the near-horizon spinor is fixed by the behavior at $x=0$.

\subsection{Less-singular near-horizon branch}
\label{subsec:near_horizon_regular_branch}

The horizon corresponds to $x=0$. Since the local Heun function is finite at the regular singular point and $e^{\tau_{\rm H}x}\to1$, Eq.~\eqref{eq:heun_factorization} gives
\begin{equation}
    \psi_1(\rho)
    \sim
    x^{q_+} Z(0),
    \qquad
    x\to0.
    \label{eq:psi1_near_horizon_x}
\end{equation}
Thus, the Heun factorization reproduces directly the less-singular $q_+$ branch found from the indicial equation in Eq.~\eqref{eq:qpm_p1}.

The local solution is normalized by choosing
\begin{equation}
    Z(0)=1.
    \label{eq:Z0_normalization}
\end{equation}
With this convention,
\begin{equation}
    \psi_1(\rho)
    \sim
    \left(
        -\frac{m\rho}{\Omega}
    \right)^{q_+},
    \qquad
    \rho\to0^+.
    \label{eq:psi1_near_horizon_rho}
\end{equation}
The overall multiplicative constant is arbitrary because the radial Dirac system is linear. It is fixed only when a normalization prescription or a finite-radius matching condition is imposed.

The second independent exponent is the $q_-$ branch of Eq.~\eqref{eq:qpm_p1}. The local solution used in this work is built from $q_+$, according to the branch convention specified there.

The less-singular branch specifies local boundedness in the chosen corotating near-horizon orthonormal frame. In the undeformed limit, the radial factors become $\rho^{-1/2\pm i\Omega}$; their ingoing and outgoing character is defined instead in horizon-regular Eddington--Finkelstein coordinates with a regular spin frame. The singular interactions also modify the leading eigenspinors. Accordingly, the branch selected here supplies local near-horizon data for the finite-radius response, while a quasinormal prescription requires a separate horizon-regular construction.

\subsection{Reconstruction of the second spinor component}
\label{subsec:spinor_reconstruction}

The confluent-Heun reduction determines the first spinor component. The second component is not independent and must be reconstructed from the original first-order radial system. Combining Eq.~\eqref{eq:psi2_from_psi1_p1} with the denominator identity \eqref{eq:epsilon_plus_m_p1} gives
\begin{equation}
    \psi_2(\rho)
    =
    -
    \frac{\rho}
    {\Omega+m\rho}
    \left\{
        \psi_1'(\rho)
        +
        \left[
            \lambda_0
            +
            \frac{\beta_1+\frac{1}{2}}{\rho}
        \right]\psi_1(\rho)
    \right\}.
    \label{eq:psi2_reconstruction_compact}
\end{equation}

Substituting the factorization \eqref{eq:heun_factorization} and the derivative transformation \eqref{eq:rho_x_derivatives} into Eq.~\eqref{eq:psi2_reconstruction_compact} gives the second component directly in terms of the Heun function:
\begin{align}
    \psi_2(\rho)
    =
    \frac{\rho}
    {\Omega+m\rho}
    x^{q_+} e^{\tau_{\rm H}x}
    \Bigg\{
    &
    \frac{m}{\Omega}
    \left[
        Z'(x)
        +
        \left(
            \tau_{\rm H}+\frac{q_+}{x}
        \right)Z(x)
    \right]
    \nonumber\\
    &
    -
    \left[
        \lambda_0
        +
        \frac{\beta_1+\frac{1}{2}}{\rho}
    \right]Z(x)
    \Bigg\}.
    \label{eq:psi2_heun_full}
\end{align}
Equations~\eqref{eq:heun_factorization} (with $y=\psi_1$) and \eqref{eq:psi2_heun_full} define the complete spinor solution in the componentwise Heun representation.

The near-horizon ratio between the spinor components follows directly from the leading power behavior. Taking
\begin{equation}
    \psi_1(\rho)
    \sim
    \rho^{q_+},
    \qquad
    \psi_2(\rho)
    \sim
    C\,\rho^{q_+},
    \label{eq:near_horizon_ratio_ansatz}
\end{equation}
and substituting into the leading terms of the coupled system gives
\begin{equation}
    \frac{\psi_2}{\psi_1}
    \longrightarrow
    -
    \frac{
        q_+
        +
        \beta_1
        +
        \frac{1}{2}
    }
    {\Omega},
    \qquad
    \rho\to0^+.
    \label{eq:near_horizon_spinor_ratio}
\end{equation}
This relation gives the local projective spinor datum associated with the selected near-horizon branch.

The construction shows that the Heun solution is not a scalar replacement for the Dirac problem. It is a representation of one spinor component, while the second component is fixed by the original first-order equation. The complete radial mode is therefore the spinor pair
\begin{equation}
    \Psi_{\rm rad}(\rho)
    =
    \begin{pmatrix}
        \psi_1(\rho)\\
        \psi_2(\rho)
    \end{pmatrix},
    \label{eq:radial_spinor_pair}
\end{equation}
and equivalence between alternative representations must be understood at the level of this complete pair.

\subsection{Analytical equivalence of the Heun and Whittaker representations}
\label{subsec:heun_whittaker_analytic_equivalence}

The componentwise confluent-Heun form and the Whittaker form are not independent solutions of different radial problems. Both are exact representations of the same first-order $p=1$ spinor system. The Heun equation is obtained by eliminating $\psi_2$ from Eqs.~\eqref{eq:p1_system_1}--\eqref{eq:p1_system_2}, while Eq.~\eqref{eq:psi2_reconstruction_compact} reconstructs the eliminated component. Therefore the pair
\begin{equation}
    \Psi_{\rm H}(\rho)
    =
    \begin{pmatrix}
        \psi_1^{\rm H}(\rho)\\
        \psi_2^{\rm H}(\rho)
    \end{pmatrix}
    \label{eq:heun_spinor_pair_analytic}
\end{equation}
satisfies the original coupled system identically on every interval where the reconstruction denominator $\epsilon(\rho)+m$ is nonzero.

The Whittaker representation follows instead from the constant spinor transformation of Sec.~\ref{subsec:whittaker_reduction}. If $\Phi_{\rm W}$ denotes a solution of the transformed system, then
\begin{equation}
    \Psi_{\rm W}(\rho)
    =
    T\Phi_{\rm W}(\rho)
    \label{eq:whittaker_spinor_pair_analytic}
\end{equation}
satisfies the same original first-order equations because $T$ is constant and invertible in the generic case. Hence, once the same local branch and projective datum are chosen, uniqueness of the first-order system implies
\begin{equation}
    \Psi_{\rm H}(\rho)
    =
    C\,\Psi_{\rm W}(\rho),
    \label{eq:heun_whittaker_equivalence_analytic}
\end{equation}
with a $\rho$-independent normalization constant $C$. In particular,
\begin{equation}
    \frac{\psi_2^{\rm H}}{\psi_1^{\rm H}}
    =
    \frac{\psi_2^{\rm W}}{\psi_1^{\rm W}},
    \label{eq:heun_whittaker_response_equivalence}
\end{equation}
wherever the ratios are defined.

This also clarifies the role of the finite point $\rho=-\Omega/m$ appearing in the componentwise scalar equation. The pole is introduced by dividing by $\epsilon(\rho)+m$ during elimination; it is absent from the original matrix-Coulomb system and from the rotated Whittaker system. The additional finite singularity is therefore apparent at the level of the complete spinor rather than a new singularity of the underlying first-order dynamics.

\subsection{Constant-coupling limit}
\label{subsec:usual_oscillator_limit}

The constant-coupling near-horizon Dirac oscillator of Ref.~\cite{guvendi2024} is recovered by setting $\beta_1\to0$ in Eq.~\eqref{eq:lambda_iso_definition}. The coupling then reduces to $\mathcal{L}_1(\rho)=\lambda_0$, and Eqs.~\eqref{eq:p1_system_1}--\eqref{eq:p1_system_2} reduce directly to the reference constant-coupling radial system. In that reference case, setting $\lambda_{\rm osc}=m\omega_0/\kappa$ in Eq.~\eqref{eq:lambda0_definition} identifies $\lambda_0$ with the constant $\lambda$ of Eq.~\eqref{eq:usual_lambda}.

The componentwise decoupling remains sensitive to the radial denominator in Eq.~\eqref{eq:epsilon_plus_m_p1}; hence the apparent point $\Omega+m\rho=0$ persists in the scalar equation even when $\beta_1=0$, while the complete first-order system remains regular there.

For generic $\Omega\neq0$, Eq.~\eqref{eq:qpm_p1} gives $\sigma=\sqrt{-\Omega^2}$ and Eq.~\eqref{eq:whittaker_variables} gives $\kappa_{\rm W}=-1/2$. The exact Whittaker solution therefore reduces continuously to
$\rho^{-1}M_{-1/2,\,\pm\sqrt{-\Omega^2}}(2\Xi\rho)$ in the rotated basis. The componentwise Heun representation has the same continuous limit. Thus, the $p=1$ model is a continuous one-parameter extension of the reference constant-coupling system, and the matrix formulation makes explicit that the generic limit belongs to the Whittaker class.

\end{document}